\documentclass[%
 reprint,
superscriptaddress,
 amsmath,amssymb,
 aps,
]{revtex4-2}

\usepackage{graphicx}
\usepackage{dcolumn}
\usepackage{bm}
\usepackage{hyperref}
\hypersetup{colorlinks=true, linkcolor=blue, anchorcolor=black, citecolor=blue,
urlcolor = blue}
\usepackage{comment}
\usepackage[hyphenbreaks]{breakurl}
\usepackage{xcolor}

\begin{document}
\newcommand{\scale}{.9}
\preprint{APS/123-QED}

\title{Rydberg atom reception of a handheld UHF frequency-modulated two-way radio}

\author{Noah Schlossberger}
 \email{noah.schlossberger@nist.gov}
\affiliation{National Institute of Standards and Technology, Boulder, Colorado 80305, USA}
\author{Tate McDonald}
\affiliation{Department of Physics, University of Colorado, Boulder, Colorado 80309, USA}
\affiliation{National Institute of Standards and Technology, Boulder, Colorado 80305, USA}
\author{Nikunjkumar Prajapati}%
\affiliation{National Institute of Standards and Technology, Boulder, Colorado 80305, USA}

\author{Christopher L. Holloway}
\affiliation{National Institute of Standards and Technology, Boulder, Colorado 80305, USA}

\date{\today}

\begin{abstract}
Rydberg atoms, due to their large polarizabilities and strong transition dipole moments, have been utilized as sensitive electric field sensors. While their capability to detect modulated signals has been previously demonstrated, these studies have largely been limited to laboratory-generated signals tailored specifically for atomic detection. Here, we extend the practical applicability of Rydberg sensors by demonstrating the reception of real-world frequency-modulated (FM) audio transmissions using a consumer-grade handheld two-way radio operating in the UHF band. Detection is based on the AC Stark shift induced by the radio signal in a Rydberg atomic vapor, with demodulation performed using an offset local oscillator and lock-in amplification. We successfully demodulate speech signals and evaluate the audio spectral response and reception range. We show that all consumer-accessible radio channels can be simultaneously detected, and demonstrate simultaneous reception of two neighboring channels with at least 53~dB of isolation. This work underscores the potential of Rydberg atom-based receivers for practical, real-world FM signal detection.
\end{abstract}
\maketitle


\section{Introduction}
Alkali atoms in highly excited Rydberg states make sensitive electric field sensors due to their large transition dipole moments and polarizabilities \cite{Schlossberger2024Nature}. These quantum field sensors are powerful tools with applications in imaging \cite{Fan:14,CardmanGonçalvesSapiroRaithelAnderson+2020+305+312, schlossberger2024twodimensionalimagingelectromagneticfields, behary2025electronbeamcharacterizationfluorescence} and angle-of-arrival determination \cite{10.1063/5.0045601, 10.1063/5.0240787, 11000341, talashila2025determininganglearrivalradio, 10304615, gill2025microwavephasemappingangleofarrival, 6dl6-754w, pthj-gy98} of RF fields, radar \cite{chen2025highresolutionquantumsensingrydberg, watterson2025imagingradarusingrydberg}, and blackbody radiometry \cite{PhysRevApplied.23.044037,PhysRevResearch.7.L012020, Borówka2024}. One of the largest application spaces for Rydberg sensors is communications. Rydberg atoms are capable of receiving both AM and FM modulated signals \cite{HollowayAMFM}, allowing them to receive signals such as audio \cite{10.1063/1.5099036} and analog video \cite{10.1116/5.0098057}. However, most of these demonstrations have involved detecting a signal that was curated for the Rydberg receiver, with a carrier frequency and modulation scheme chosen to match the capabilites of the sensor. Only a limited body of work \cite{lei2025satellitesignaldetectionrydbergatom, 10.1063/5.0158150} has examined the use of these atomic sensors to receive real-world signals operating on existing protocols, and these demonstrations have involved collecting the radiation with a classical antenna and focusing it down onto the atomic sensor.

Here we demonstrate that without any focusing or field enhancement, Rydberg atoms can detect the audio signal from a consumer-grade handheld two-way radio, or ``walkie talkie.'' As these radios transmit in the ultra-high frequency (UHF) band, we detect their signal by the AC Stark shift they induce on a Rydberg state. This technique is very general as the detection scheme is independent of the carrier frequency. We demonstrate the reception of handheld radio signals with modest postprocessing (only a lock-in amplifier operating at 100~kHz). We also demonstrate that we can simultaneously receive all 22 radio channels, requiring only a separate lock-in filter to decode each channel. We demonstrate this idea by simultaneously and independently detecting two neighboring channels.

\section{FRS radio}
\begin{figure}
\raggedright
\textbf{(a)}\\
\centering
\scriptsize
\begin{tabular}{|c|c||c|c|}
\hline
channel & frequency &channel & frequency \\ \hline
1 & 462.5625 MHz & \textcolor{gray}{12} & \textcolor{gray}{467.6625 MHz} \\ \hline
2 & 462.5875 MHz & \textcolor{gray}{13} & \textcolor{gray}{467.6875 MHz} \\ \hline
3 & 462.6125 MHz & \textcolor{gray}{14} & \textcolor{gray}{467.7125 MHz} \\ \hline
4 & 462.6375 MHz & 15 & 462.5500 MHz \\ \hline
5 & 462.6625 MHz & 16 & 462.5750 MHz \\ \hline
6 & 462.6875 MHz & 17 & 462.6000 MHz \\ \hline
7 & 462.7125 MHz & 18 & 462.6250 MHz \\ \hline
\textcolor{gray}{8} & \textcolor{gray}{467.5625 MHz} & 19 & 462.6500 MHz \\ \hline
\textcolor{gray}{9} & \textcolor{gray}{467.5875 MHz} & 20 & 462.6750 MHz \\ \hline
\textcolor{gray}{10} & \textcolor{gray}{467.6125 MHz} & 21 & 462.7000 MHz \\ \hline
\textcolor{gray}{11} & \textcolor{gray}{467.6375 MHz} & 22 & 462.7250 MHz \\ \hline
\end{tabular}
\normalsize
\\
\raggedright
\textbf{(b)}\\
\includegraphics[scale = \scale]{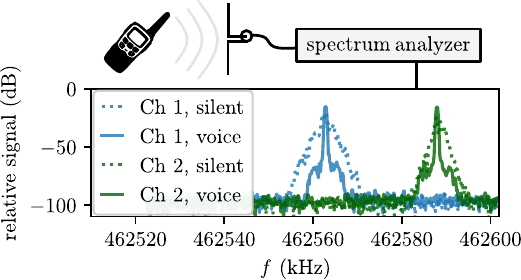}
    \caption{FRS radio channels. a) A table of center frequencies of all 22 FRS radio channels. b) Measured spectrum analyzer traces of the handheld radio transmitting on FRS channels 1 and 2, recorded with a resolution bandwidth of 300 Hz.}
    \label{fig:FRS}
\end{figure}

RF audio transmission in the UHF band without an operating license in the United States is limited to the Family Radio Service (FRS), which includes the 462.550–462.725~MHz band with an output power limited to 2~W, and the 467.5625–467.7125~MHz band with an output power limited to 0.5~W \cite{FCCregulations}. These frequency bands are divided into 22 separate channels. The carrier frequencies of each channel are given in Fig.~\ref{fig:FRS}a. 
The audio is then encoded in a frequency modulation of the carrier. The maximum FM depth, or deviation of the instantaneous frequency from the carrier, is 2.5 kHz, and the maximum FM bandwidth, or encoded audio frequency, is 12.5 kHz.

In Fig.~\ref{fig:FRS}b, we measure the broadcast of FRS channels 1 and 2 with an unmatched, electrically small dipole antenna feeding into a spectrum analyzer. We transmit using a Midland T51 \cite{NISTDisclaimer} two-way radio. When the radio operator is silent, no FM is applied and the signal appears narrowband. When the radio operator speaks, the signal is spread over the bandwidth of the channel. Note that although the signal appears to be many kHz wide, the instantaneous frequency is deviating by less than 2.5~kHz (the resolution bandwidth is slow compared to the audio signal).

\section{Measurement scheme}
\begin{figure}
    \includegraphics[scale = \scale]{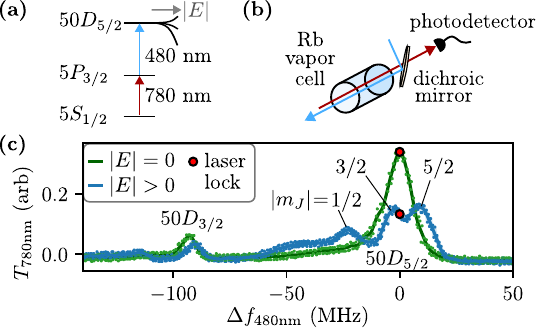}
    \caption{Measurement of the electric field strength using Rydberg atoms. a) The energy level diagram. b) The physical setup. c) The measured EIT spectrum, given by the transmission of the 780~nm light ($T_\textrm{780nm}$) as a function of the detuning of the 480~nm light ($\Delta f_\textrm{480nm}$).}
    \label{fig:eld}
\end{figure}

To detect FRS radio signals, we use $^{85}$Rb atoms as our atomic sensor. We measure the energy of a Rydberg state ($50D_{5/2}$) using a conventional \cite{Finkelstein_2023, PhysRevLett.98.113003} two-photon ladder scheme to create an electromagnetically induced transparency (EIT). The energy level diagram is shown in Fig.~\ref{fig:eld}a. The two beams are counter-propagating (Fig.~\ref{fig:eld}b) in order to partially cancel the Doppler shift due to thermal motion of the atoms, resulting in narrower spectral features.  The 780~nm laser has a beam full width at half max (FWHM) of 610~$\mu$m and a power of 140~$\mu$W, while the 480~nm laser has a beam FWHM of 890~$\mu$m and a power of 490~mW.

The presence of a UHF band RF field will induce an AC Stark shift on the Rydberg state. To first order, the energy of the Rydberg state is shifted by \cite{Delone:1999}
\begin{equation}
   \Delta U_\textrm{Stark}= \frac{1}{2} \alpha_{|m_J|}(f) E^2,
\end{equation}
where $\alpha(f)_{|mJ|}$ is the AC polarizability of each fine structure projection $m_J$ state, $f$ is the frequency of the RF field, and $E$ is the root-mean-square magnitude of the applied electric field. Because each magnitude $|m_J|$ has a different polarizability $\alpha_{|m_J|}$, the RF field will split the spectra into three peaks (Fig.~\ref{fig:eld}c).
When we lock the lasers on resonance, a change in the magnitude of the UHF field is mapped into a change in transmission of the 780~nm light (red dots in Fig.~\ref{fig:eld}c), and thus a change in the photodetector voltage. 

With this scheme, the photodetector voltage tracks the magnitude of the applied field. However, this scheme is insensitive to the frequency of the applied field. The polarizabilities are constant to within $10^{-6}$ of a percent over the span of one FRS channel \cite{ROBERTSON2021107814}. This means that the frequency modulation does not affect the EIT spectrum. In order to detect the FM encoded audio, we apply a local oscillator (LO) with an unmatched pair of wires around the vapor cell. Because the electrodes are spaced by much less than a wavelength of the UHF radiation, this creates a localized RF field inside of the vapor cell. 

\begin{figure}
    \includegraphics[scale = \scale]{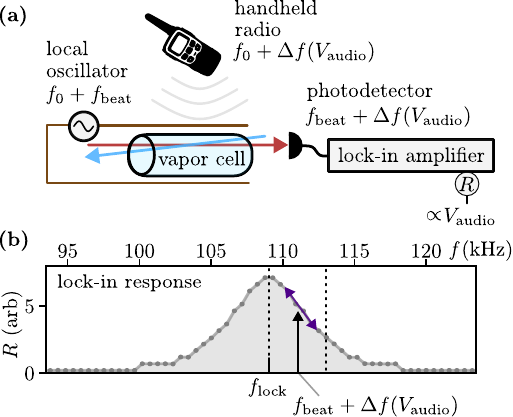}
    \caption{Demodulation scheme for receiving FM signals via AC Stark shift. a) The detection scheme. The frequency is denoted at each point along the signal chain to demonstrate the encoding of the audio signal. b) The lock-in amplifier's measured amplitude ($R$) response to various frequencies with the lock-in frequency $f_\textrm{lock}$ set to 108.5~kHz. The slope of the response is used to convert the FM modulation of the beatnote to a voltage.}
    \label{fig:demodscheme}
\end{figure}

The audio detection scheme is shown in Fig.~\ref{fig:demodscheme}.
The handheld radio transmits at a center frequency $f_0$ (as in Fig.~\ref{fig:FRS}a), with a frequency modulation proportional the audio signal $\Delta f (V_\textrm{audio})$. We then apply a constant tone with a local oscillator a frequency $f_\textrm{beat}$ away from the carrier frequency. The two fields interfere inside of the vapor cell, resulting in a beatnote between the two fields. This causes the amplitude of the UHF field inside of the cell to vary at a beat frequency of 
\begin{equation}
    |(f_0 + f_\textrm{beat}) - (f_0 + \Delta f(V_\textrm{audio}))| = f_\textrm{beat} + \Delta f(V_\textrm{audio}).
\end{equation}
This amplitude variation is then mapped to the transmission of the laser light, meaning the photodetector voltage will oscillate at this frequency. With this, we have mapped the FM of the radio at a carrier $f_0$ to FM on the voltage of a photodetector at a carrier $f_\textrm{beat}$. 

Finally, to convert the FM of the beatnote into an audio signal, we use a lock-in amplifier. We select a lock-in frequency $f_\textrm{lock}$ slightly different to the beat frequency $f_\textrm{beat}$, so that the signal is sitting on the side of the lock-in amplifier's response. The response of the lock-in detector then turns the frequency modulation into a modulation on the lock-in amplitude $R$, which we can detect as a voltage from the lock-in amplifier's $R$ port. This can then be directly connected to speakers or an audio interface.

\section{Single channel detection}
\begin{figure}
. \includegraphics[scale = \scale]{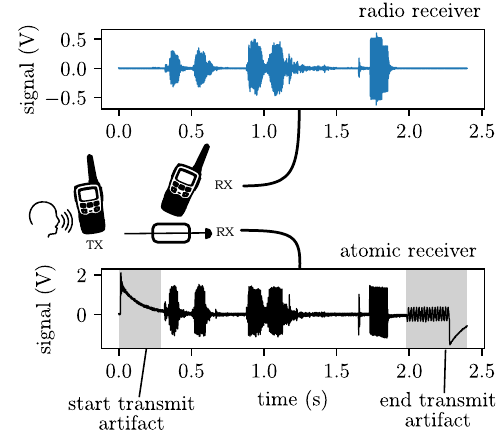}
   \caption{Simultaneous recordings of a voice speaking into a handheld two-way radio, recorded using the line out of another handheld two-way radio (top) and the atomic receiver (bottom). 
   }
   \label{fig:receive_singlech}
\end{figure}

We demonstrate this scheme by detecting FRS channel 1, with an $f_0$ of 462.5625 MHz. We choose a beatnote that is fast compared to the audio signal ($\sim$kHz) such that it can be demodulated, but slow compared to the atomic bandwidth ($\sim$10~MHz \cite{10.1063/1.5028357, Hu2023, PhysRevApplied.21.L031003, PhysRevA.111.033115}). We therefore choose a beatnote frequency $f_\textrm{beat} = 111$~kHz, corresponding to an LO frequency of 462.4515~MHz. We then set the data transfer rate of the lock-in amplifier (related to the width of the amplitude response) to 1~kHz, and set $f_\textrm{lock}$ to 108.5~kHz (2.5~kHz below $f_\textrm{beat}$). After connecting the amplitude port of the lock-in amplifier to an audio interface, we can record human speech transmitted over the two-way radio (Fig.~\ref{fig:receive_singlech}).

The artifacts that occur at the beginning and end of the transmission are due to the turning on and off of the carrier. With no carrier, the value of $R$ will be zero. When the broadcast begins, the carrier generates a finite $R$, with the audio-induced FM changing the response of $R$. This essentially creates a DC bias to the signal, which is removed via AC coupling to the audio interface. When the broadcast begins and ends, this DC bias is suddenly added or removed, which passes throught the AC coupling into the audio interface.

\begin{figure}
\includegraphics[scale = \scale]{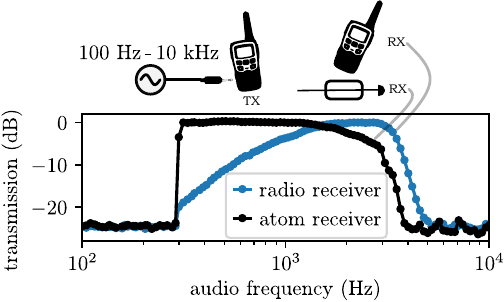}
\caption{Spectral response of audio transmitted by a handheld two-way radio and received by either another handheld two-way radio or the atomic receiver.  The spectra are normalized to the largest transmitted power to account for different electronic gains of the two receivers.}
\label{fig:audio_spectrum}
\end{figure}

Next, we characterize the audio spectral response of the atomic receiver. We connect a function generator to the microphone port of one of the handheld radios to transmit a pure tone, and measure the corresponding transmitted power. The results are shown in Fig.~\ref{fig:audio_spectrum}.

No transmission occurs below 300~Hz because the two-way radio transmitter does not transmit frequencies below 300~Hz. Over the audio range, the atomic receiver is relatively flat while the radio receiver seems to act as a low-pass filter, most likely due to internal post-demodulation filtering built into the receiver. The radio receiver's response falls off at around 3~kHz, while the atomic receiver falls off slightly lower at around 2~kHz because the lock-in amplifier's output has a fourth order low-pass filter with a 3~dB point at 2~kHz.

\begin{figure}
    \includegraphics[scale = \scale]{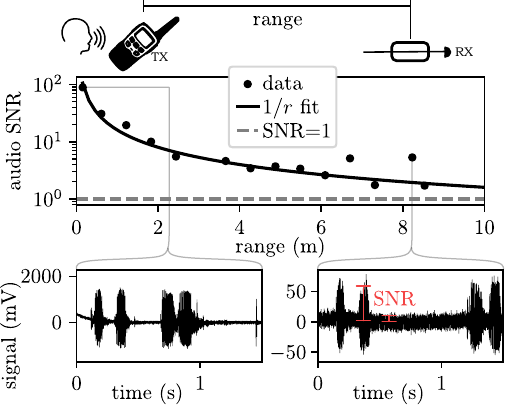}
    \caption{The signal-to-noise ratio (SNR) of the atomic receiver detecting a two-way radio transmission at varying ranges, corrected as in Eq. \ref{eq:SNRcorr}.}
    \label{fig:range}
\end{figure}

Finally, we characterize the range of the atomic radio receiver by measuring the signal to noise ratio (SNR) at various ranges (Fig.~\ref{fig:range}). The SNR is determined by taking the ratio of the RMS values of the audio signal during transmission taken during speech vs when not speaking. We then correct for the noise contributing an RMS value during speaking by taking
\begin{equation}
    \textrm{SNR}_\textrm{corrected} = \sqrt{\textrm{SNR}_\textrm{uncorrected}^2 - 1}. \label{eq:SNRcorr}
\end{equation}
Since the audio signal is encoded in the electric field, the strength of the signal will drop off as the strength of the electric field, which goes as $1/r$ where $r$ is the range. Because the detected noise primarily comes from spectroscopic noise which does not scale with range, it is essentially fixed, and so we expect the SNR to fall off as $1/r$. We find in Fig.~\ref{fig:range} that the SNR seems to go roughly as $1/r$, but has a large deviation because of reflections in the laboratory as well as variations in the orientation of the handheld radio.

The expected range $r$ to retain an SNR of at least a factor of 2 for a source of directivity $\mathcal{D}$ transmitting a power $P$ from an electrometer of sensitivity $\mathcal{S}$ measuring at a bandwidth of $\Gamma$ is given by
\begin{equation}
    \sqrt{\frac{\mathcal{D} P}{8 \pi c \epsilon_0 \Gamma \mathcal{S}^2}},
\end{equation}
where $c$ is the speed of light and $\epsilon_0$ is the permittivity of free space. 
Rydberg electrometers sensing in this regime (frequencies where the AC polarizability approaches the DC polarizability) have been demonstrated with sensitivities as low as 0.1~(mV/m)/$\sqrt{\textrm{Hz}}$ \cite{Yang:24}. Assuming the radio is transmitting at its maximal permitted power ($P=2$~W) and has a Hertzian dipole antenna ($\mathcal{D} = 1.5$), the ideal range we could expect from a Rydberg receiver sampling at $\Gamma = 1$~kHz is 2.1~km. Our Rb state has a lower polarizability than the Cs state used in \cite{Yang:24}; scaling the sensitivity by the polarizabilities of the states we expect our system to have a maximum ideal range of 40~m, in order-of-magnitude agreement with Fig. \ref{fig:range}.
\section{Multi-channel detection}
Because the 462 MHz portion of the FRS is constrained to a 175~kHz band, which is much less than the instantaneous bandwidth of the Rydberg receiver, every channel in this band will beat against our LO and produce a measurable beatnote on our detector, each at a different frequency. This means that the different channels can be detected simply by changing the frequency $f_{lock}$ of our lock-in detector. 

\begin{figure}
\includegraphics[scale = \scale]{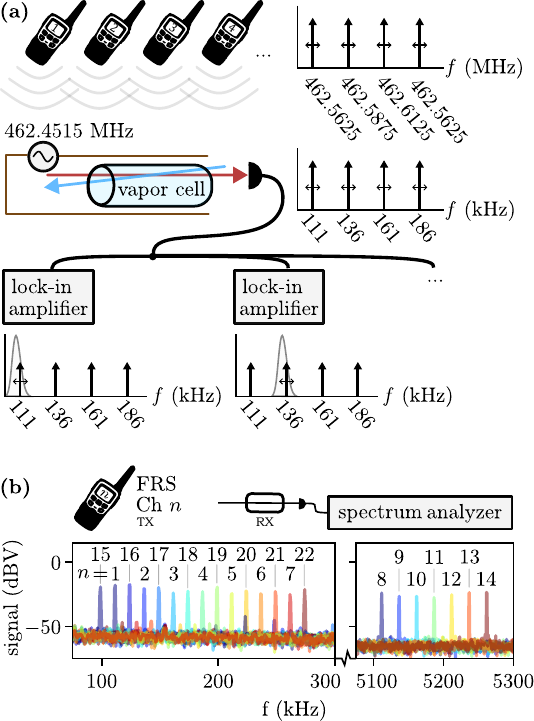}
\caption{Measurement scheme for simultaneously receiving all 462 MHz FRS channels. a) All FRS channels form independent beatnotes with the LO, and each channel is selected and recieved after the photodiode by different lock-in amplifier channels using different lock-in frequencies. b) Measured spectrum analyzer traces of the photodetector signal when broadcasting on each FRS channel with the same LO frequency of 462.4515 MHz.}
\label{fig:all_channel_scheme}
\end{figure}

In fact, we can receive multiple channels simultaneously by using multiple lock-in frequencies on our lock-in detector. A scheme of receiving and demodulating multiple FRS channels simultaneously is shown in Fig.~\ref{fig:all_channel_scheme}a.

In this scheme, a single LO will automatically generate a separate beatnote for each FRS radio channel, which can be measured by tuning the lock-in amplifier to the channel of interest. Because these beatnotes are independent, multiple channels can be simultaneously received using different channels on the lock-in amplifier. The frequency of the local oscillator relative to the first channel is chosen to be incommensurate with the channel spacing such that harmonics of the beatnotes do not interfere.

To demonstrate the reception of frequency-separated beatnotes for each FRS channel, we measure the photodetector voltage with a spectrum analyzer and record the beatnote signal as we tune the handheld radio to each FRS channel (Fig.~\ref{fig:all_channel_scheme}b). We measure beatnotes in the few hundred kHz range for FRS channels 1-7 and 15-22, which broadcast at 462~MHz. The other channels still lie within the atomic response bandwidth: we measure beatnotes at 5~MHz for channels 8-14, which broadcast at 467~MHz. These could also be demodulated given a high-bandwidth lock-in amplifier.

\begin{figure}
    \includegraphics[scale = \scale]{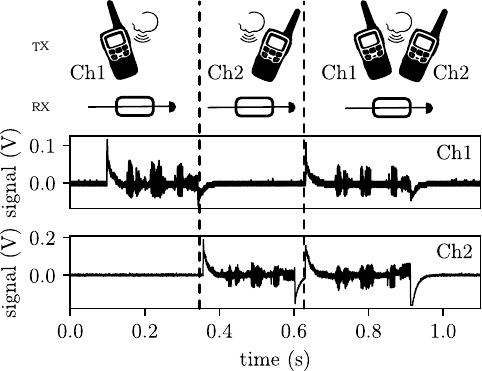}
    \caption{Simultaneous atomic reception of two FRS channels using two lock-in frequencies on the same photodetector line, with the channel 1 demodulation shown on top and channel 2 on bottom. Three voice transmissions are made, with only channel 1 transmitting (left), only channel 2 transmitting (middle), and both transmitting (right).}
    \label{fig:twochanns}
\end{figure}

We demonstrate the simultaneous detection scheme using two handheld radio transmitters and two lock-in amplifier channels. To detect channel 1 ($f_0 = 462.5625$~MHz) and channel 2 ($f_0 = 462.5875$~MHz), we set our local oscillator frequency to 462.4515~ MHz and set our lock-in frequencies to 108.5~kHz for FRS channel 1 ($f_\textrm{beat} =111$~kHz) and 133.5~kHz for FRS channel 2 ($f_\textrm{beat} = 136$~kHz). Both channels use a time constant of 900 samples/s. The recorded audio is shown in Fig.~\ref{fig:twochanns}. 

\begin{figure}
    \includegraphics[scale = \scale]{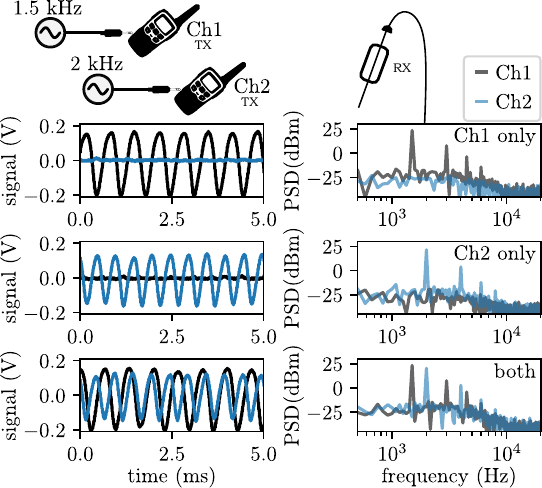}
    \caption{Measurement of channel isolation during simultaneous detection. The time trace (left) and power spectral density (PSD) (right) are shown when broadcasting a 1.5~kHz pure tone on channel 1 (top), a 2~kHz pure tone on channel 2 (middle), and both simultaneously (bottom).}
    \label{fig:isolation}
\end{figure}

Here, the two traces represent the two lock-in demodulations. It is qualitatively clear that the two channels are independently received. However, we quantitatively determine the independence of the two channels by broadcasting two separate pure tones (Fig.~\ref{fig:isolation}), which we can then separate after reception in Fourier space.

If the channels contained crosstalk, we would detect frequency components of each broadcast in the opposite channel, evident in the power spectral density. We find no crosstalk discernible from the noise floor. As such, the noise floor sets a lower bound on the isolation between channels at 53~dB of isolation from channel 1 into 2 and 40~dB of isolation from channel 2 into channel 1.

\section{Conclusion}
We have demonstrated reception of FM modulated signals using AC Stark shift in Rydberg states of alkali atoms. We successfully demodulated audio from a handheld consumer two-way radio. We characterized its range and spectral properties. We then demonstrated that we can simultaneously demodulate multiple radio channels, with at least 53~dB of isolation.

While the range we demonstrated was modest, it could be significantly improved with a more sensitive Rydberg state. Cs generally has higher polarizabilities than Rb, increasing further with higher principal quantum numbers. Three photon systems make Rydberg $F$ states accessible, which have even higher polarizabilites and resonant transitions in the UHF band \cite{PhysRevA.107.052605, 10.1063/5.0179496}. 

A major drawback of this technique is the requirement of a local oscillator. This could be removed if the Rydberg sensor could be made frequency-dependent. One could imagine placing the vapor cell in a resonant structure such that the amplitude of the field became frequency dependent. However, the width of the resonance would need to be commensurate with the 2.5~kHz FM modulation depth, requiring a quality factor on the order of $10^4$. 

Instead, the practical approach forward would be to choose Rydberg states whose response is frequency dependent in this range. Three photon schemes probing $F$-states can have much lower resonance frequencies than the $D$-states. For example, the $56F_{7/2} \leftrightarrow 56G_{9/2}$ transition in $^{85}$Rb or the $72F_{7/2}\leftrightarrow 72G_{9/d2}$ transition in Cs have resonance frequencies of 469.5~MHz and 464.8~MHz respectively, offering respective polarizability gradients of 0.03\% and 0.1\% over an FRS channel. However, because the DC polarizabilities of these states are also high, we could introduce a field gradients in the vapor cell by shining visible light to induce the photoelectric effect in alkali atoms adsorbed on the surface \cite{10.1116/5.0264378}, which would bring part of the beam into resonance with these transitions. This resonant detection would both be more sensitive and have a much higher frequency dependence, offering a promising path forward to FSR radio reception without applying an RF local oscillator.


\section*{Acknowledgments}
\vspace{-.2cm}
\noindent
This research  was supported by NIST under the NIST-on-a-Chip program.  A contribution of the U.S. government, this work is not subject to copyright in the U.S. 
\\[10pt]
\vspace{.5cm}

\noindent
All of the data presented in this paper and used to support the conclusions of this article is available at \cite{MIDAS}.
\\[5pt]
The authors declare no conflict of interest.

\begin{thebibliography}{38}
\bibitem [{NIS()}]{NISTDisclaimer}%
  \BibitemOpen
  \href@noop {} {}\bibinfo {note} {Certain equipment, instruments, software, or materials are identified in this paper in order to specify the experimental procedure adequately. Such identification is not intended to imply recommendation or endorsement of any product or service by NIST, nor is it intended to imply that the materials or equipment identified are necessarily the best available for the purpose.}\BibitemShut {Stop}%
\end{thebibliography}
\providecommand{\noopsort}[1]{}\providecommand{\singleletter}[1]{#1}%

\end{document}